\documentclass[onecollarge,natbib]{svjour2}
\bibpunct{[}{]}{;}{n}{}{,} 
\smartqed 
\usepackage{graphicx}

\newcommand{\beq}{\begin{eqnarray}}
\newcommand{\eeq}{\end{eqnarray}}
\newcommand{\nneeq}{\nonumber \end{eqnarray}}

\newcommand{\nn}{\nonumber \\}
\newcommand{\es}{& = &}
\newcommand{\rs}{\, = \,}
\newcommand{\ps}{& + &}

\newcommand{\np}{\nn \ps}

\newcommand{\cH}{ {\cal H} }

\newcommand{\cG}{ {\cal G} }
\newcommand{\cT}{ {\cal T} }

\newcommand{\cU}{ {\cal U} }
\newcommand{\cL}{ {\cal L} }
\newcommand{\cP}{ {\cal P} }

\journalname{Few Body Systems}
\begin{document}
\title{ Fermion mass mixing in vacuum }
\author{ Stanis{\l}aw D. G{\l}azek }
\institute{ Institute of Theoretical Physics, 
            Faculty of Physics, 
            University of Warsaw \at
            Ho\.za 69, 00-681 Warsaw, Poland \\
              Tel.: +48-22-553 2277\\
              Fax:  +48-22-621-9475\\
              \email{stglazek@fuw.edu.pl}}
\date{Received: 29 October 2013 / Accepted: 11 November 2013}

\maketitle

\begin{abstract}
Renormalization group procedure for effective
particles (RGPEP) is applied to a theory of 
fermions that interact only through mass mixing
terms in their Hamiltonian. Problems with virtual 
pair production in vacuum are avoided by using 
the front form of Hamiltonian dynamics. Masses 
and states of physical fermions emerge at the 
end of a calculation that is carried out exactly 
irrespective of the strength of the mass mixing 
terms. An a priori infinite set of renormalization 
group equations for all momentum modes of fermion
quantum fields is reduced to just one equation 
for a two-by-two mass matrix. In distinction 
from scalars, fermions never become tachyons 
but appear chirally rotated when the mass mixing 
interaction term is sufficiently strong.
\end{abstract}

\section{Introduction}
\label{intro}

Among the issues that need to be dealt with on
the way to defining and solving relativistic 
quantum field theory beyond the perturbative 
pictures based on free propagation or weak 
binding of field quanta, one may distinguish 
the issues of vacuum and renormalization, both 
intricately related with the struggle to understand 
relativity and quantum mechanics simultaneously
in one theory. Literature concerning these two 
issues is of enormous size and scope and this
article is not to review it. Instead, this article 
offers a description of how the renormalization 
group procedure for effective particles (RGPEP) 
can be applied to an elementary quantum field 
theory for the relativistic fermions that 
interact through a mass mixing term of arbitrary 
strength~\cite{Glazek S(2013)}. The idea is to 
present how the RGPEP maintains symmetries of 
special relativity in an interacting quantum 
field theory and keeps the vacuum state simple. 

The example is solved exactly. The vacuum issue 
is addressed by using the front form (FF) of 
dynamics instead of the familiar instant form 
(IF)~\cite{Dirac P(1949)}. Therefore, the vacuum 
divergences~\cite{Dirac P(1965)} due to 
fermion-anti-fermion pair creation do not 
occur. The remaining interactions are included
in the RGPEP equations. The equations produce a 
relativistic spectrum of solutions in agreement 
with the general rules of quantum representation 
of symmetries of special relativity~\cite{Wigner E(1938)}. 

Since the calculation described here concerns 
basic issues and differs from the IF approach, 
a lot of basic details are included explicitly. 
First, the IF approach to the model example is 
reviewed, introducing useful notation in a 
familiar context and pointing out difficulties. 
Then the RGPEP approach is presented in a concise 
form that indicates new elements in comparison 
with the IF approach.

\section{ Theory of two kinds of free fermion fields }
\label{free}

The standard form of Lagrangian density for 
two kinds of free fermion fields $\psi$ and 
$\phi$ with mass parameters $\mu$ and $\nu$, 
respectively, is 
\beq
\label{LIF}
\cL 
\es 
\bar \psi (i \partial \hspace{-5.5pt}/ - \mu) \psi 
+
\bar \phi (i \partial \hspace{-5.5pt}/ - \nu) \phi \, .
\eeq
The condition of stationary action, $\delta S =0$,
where $S = \int d^4x \, \cL$, implies equations of 
motion 
$(i \partial \hspace{-5.5pt}/  - \mu)\, \psi = 0$ 
and 
$(i \partial \hspace{-5.5pt}/  - \nu)\, \phi = 0$.
The corresponding canonical Hamiltonian in 
the IF of dynamics is
\beq
\label{HIF}
H \es \int d^3x \, \cT^{00} \, ,
\eeq
where $\cT^{00}$ can be identified as 
the time-time component of the 
energy-momentum density tensor 
\beq
\cT^{\rho \sigma} 
\es
{\partial \cL \over \partial \partial_\rho
\psi_\alpha} \, \partial^\sigma \psi_\alpha
+
{\partial \cL \over \partial \partial_\rho
\phi_\alpha} \, \partial^\sigma \phi_\alpha
-g^{\rho \sigma} \, \cL \, .
\eeq
The integral in Eq.~(\ref{HIF}) extends over 
the three-dimensional volume in space-time 
that is defined by the condition $t=0$, where 
$t$ is the time co-ordinate of the inertial 
observer who defines the theory. As a result 
of direct evaluation, the IF Hamiltonian for 
fermion fields according to this observer has 
the form
\beq
\label{HIF1}
H = \int d^3x \, 
\left[ \,
\psi^\dagger (i \vec \alpha \vec \partial + \beta \mu) \psi 
+
\phi^\dagger (i \vec \alpha \vec \partial + \beta
\nu) \phi \,
\right] \, .
\eeq
The corresponding FF Hamiltonian is defined
differently. This will be explained later.

\section{ IF canonical quantum theory for two kinds of free fermions }
\label{Qfree}

In order to obtain a canonical quantum 
theory of fermions according to the IF 
rules~\cite{Heisenberg and Pauli(1929),
Heisenberg and Pauli(1930)} one writes 
the fields $\psi$ and $\phi$ as functions
of $\vec x$ at $t=0$ in terms of their 
Fourier components,
\beq
\psi(\vec x \,) 
\es 
\sum_{~~~~ \mu p s} \hspace{-19pt}\int  \, 
\left[ u_{\mu p s} \, b_{\mu p s}         \, e^{ i\vec p \, \vec x} 
     + v_{\mu p s} \, d_{\mu p s}^\dagger \, e^{-
i\vec p \, \vec x} \right] \, , \\
\phi(\vec x \,) 
\es 
\sum_{~~~~ \nu p s} \hspace{-19pt}\int  \, 
\left[ u_{\nu p s} \, b_{\nu p s}         \, e^{ i\vec p \, \vec x} 
     + v_{\nu p s} \, d_{\nu p s}^\dagger \, e^{-
i\vec p \, \vec x} \right]  \, ,
\eeq
where the integration over momentum $\vec p$ 
and sum over spins is denoted by the symbol
\beq
\sum_{~~~~ \mu p s} \hspace{-19pt}\int 
\es
\sum_{s = \, \pm 1} \int { d^3 p \over
(2\pi)^3 2 E_{\mu p} } 
\eeq
and the energy is $E_{\mu p} \rs \sqrt{ \mu^2 + 
\vec p^{\,\, 2} }$. Similar definitions apply 
to the field $\phi$  with mass $\nu$ in place 
of $\mu$. The coefficients $u$ and $v$ in the 
Fourier expansions are spinors of freely moving 
fermions. 

It is important for the discussion 
that follows to realize that these spinors are 
obtained by boosting
spinors~\cite{Bjorken and Drell(1964)} corresponding to fermions at rest,
$u_{\mu 0 s}$ and $v_{\mu 0 s}$ normalized to 
$\sqrt{2\mu}$ and $\sqrt{2\nu}$, respectively.
The explicit formulae are
\beq
\label{umupsIF}
u_{\mu p s} \es B(\mu, \vec p\,) \, u_{\mu 0 s} 
\quad {\rm and } \quad \quad 
v_{\mu p s} \rs B(\mu, \vec p\,) \, v_{\mu 0 s} \, ,
\eeq
where the matrix $B$ represents the boosts,
\beq
B(\mu, \vec p\,) \es {1  \over \sqrt{ 2 \mu 
(E_{\mu p} + \mu ) } } \, 
\left( p \hspace{-4.1pt}/ \beta + \mu \right) \, .
\eeq
Similar definitions hold for fermions with mass 
$\nu$. This observation is important because the 
boosting requires knowledge of the fermion mass 
and energy. The energy is treated in the boost 
matrices $B$ as if the fermions were free. Therefore, 
one may expect difficulties to occur when a theory 
includes the interactions that influence the value 
of mass and alter the expression for energy. The
difficulties should be expected due to any form
of energy that significantly differs from the free 
one. The usage of spinors of freely moving fermions 
for construction of strongly interacting quantum 
field operators is unlikely to be legitimate. 

In the absence of interactions, having the Fourier 
expansions in place, one turns the fields $\psi$ 
and $\phi$ into quantum field operators $\hat \psi$ 
and $\hat \phi$ by introducing the anti-commutation 
relations~\cite{Bjorken and Drell(1965),Weinberg S(1995),
Peskin M(1995)} 
\beq
\label{acrpsi}
\left\{ \hat \psi(\vec x\,), \hat \psi^\dagger(\vec x\, ') \right\}
\es 
\left\{ \hat \phi(\vec x\,), \hat \phi^\dagger(\vec x\, ') \right\}
\rs
\delta^3(\vec x - \vec x \, ') \, .
\eeq
These are satisfied as a result of imposing anti-commutation 
relations on the coefficients $b$ and $d$ in the Fourier expansions, 
turning them into creation and annihilation operators. Namely,  
one sets
\beq
\label{acrbd}
\left\{ b_{\mu p s}, b^\dagger_{\mu p' s'} \right\}
\es
\left\{ d_{\mu p s}, d^\dagger_{\mu p' s'}
\right\} 
\rs
2E_{\mu p} (2\pi)^3 \delta^3( \vec p - \vec p \,
') \, \delta_{s s'} \, ,
\eeq
for fermions of mass $\mu$, and similar relations for 
fermions of mass $\nu$. Normal-ordered quantum Hamiltonian,
\beq
\label{HIFQ1}
\hat H_0 = \int d^3x \, 
{:} \left[ \,
\hat \psi^\dagger (i \vec \alpha \vec \partial + \beta \mu) \hat \psi 
+
\hat \phi^\dagger (i \vec \alpha \vec \partial + \beta
\nu) \hat \phi \,
\right]{:} 
\eeq
is then obtained in the physically right form 
for free fermions, i.e., 
\beq
\label{HIFQ2}
\hat H_0 \es \hspace{-6pt}
\sum_{~~~~ \mu p s} \hspace{-18pt}\int 
\,
E_{\mu p} \,
\left( b^\dagger_{\mu p s}  b_{\mu p s}
     + d^\dagger_{\mu p s}  d_{\mu p s} \right)
+
\hspace{-6pt}
\sum_{~~~~ \nu p s} \hspace{-18pt}\int 
\,
E_{\nu p} \,
\left( b^\dagger_{\nu p s}  b_{\nu p s}
     + d^\dagger_{\nu p s}  d_{\nu p s} \right) \, .
\eeq

\section{ IF canonical quantum theory for
          fermions interacting through mass mixing }
\label{Qint}

In the IF approach, one adds interaction
terms to the free Lagrangian density. In 
the case of mass mixing terms, starting
from  
\beq
\label{LIFI}
\cL 
\es 
\bar \psi (i \partial \hspace{-5.5pt}/ - \mu) \psi 
+
\bar \phi (i \partial \hspace{-5.5pt}/ - \nu) \phi
- m \left( \bar \psi \phi + \bar \phi \psi \right)
\, ,
\eeq
and following the same construction steps for 
quantum field operators as above, one obtains 
the corresponding canonical quantum Hamiltonian, 
$\hat H = \hat H_0 + \hat H_I $, in which 
\beq
\hat H_I \es m \int d^3x \, : \left( \hat \psi^\dagger \gamma^0  \hat \phi 
         + \hat \phi^\dagger \gamma^0  \hat \psi
\right) : \, .
\eeq
Having defined the quantum field operators as in
the free theory, one arrives at
\beq
\label{HIIF}
\hat H_I 
\es
m 
\sum_{~~~~ \mu p  s} \hspace{-18pt}\int \,\,\,  
\sum_{s'} {1 \over 2 E_{\nu p} }
\left[
\bar u_{\mu p  s } \, u_{\nu p  s'} \, 
b_{\mu p  s }^\dagger \, b_{\nu p  s'} 
+
\bar u_{\mu p  s } \, v_{\nu -p s'} \, 
b_{\mu p  s }^\dagger \, d_{\nu -p s'}^\dagger
\right.
\np
\left.
\bar v_{\mu p  s } \, u_{\nu -p s'} \, 
d_{\mu p  s }         \, b_{\nu -p s'}  
-
\bar v_{\mu p  s } \, v_{\nu p s'} \, 
d_{\nu p s'}^\dagger \, d_{\mu p  s } 
\right]
+
(\mu \leftrightarrow \nu) \, .
\eeq
Physically, this result is not acceptable~\cite{Dirac P(1965)}. 
The interaction creates a badly divergent vacuum
problem. To see the problem, consider the second
term in $\hat H_I$ in Eq.~(\ref{HIIF}),
\beq
\label{hIF}
\hat h
\es 
m 
\sum_{~~~~ \mu p  s} \hspace{-18pt}\int \,\,\,  \sum_{s'} {1 \over 2 E_{\nu p} }
\,
\bar u_{\mu p  s } \, v_{\nu -p s'} \, 
b_{\mu p  s }^\dagger \, d_{\nu -p s'}^\dagger \, , 
\eeq
where the spinor matrix element is
\beq
\label{mel}
\bar u_{\mu p  s } \, v_{\nu -p s'} 
\es
\left(
\sqrt{ E_{\nu p} + \nu \over  E_{\mu p} + \mu }
+
\sqrt{ E_{\mu p} + \mu \over  E_{\nu p} + \nu }
\, \right)
\,
\chi_s^\dagger \, \vec \sigma \vec p \, i \sigma^2
\, \chi_{s'} \, ,
\eeq
and ask what the norm of state $|h\rangle = \hat h 
|0\rangle$ is. Direct evaluation yields 
\beq
\langle h | h \rangle 
\es
\langle 0 | \hat h^\dagger \hat h |0 \rangle 
\rs
V m^2 
\sum_{~~~~ \mu p  s} \hspace{-18pt}\int \,\,\,  
\sum_{s'} {1 \over 2 E_{\nu p} }
| \bar u_{\mu p  s } \, v_{\nu -p s'} |^2 
\ \sim \
V \, 4m^2 
\int { d^3 p \over (2\pi)^3 2 E_{\mu p} } \, 
{\vec p\,^2 \over E_{\nu p} } \, ,
\eeq
where $V$ denotes the volume of space at $t=0$ 
and the approximation sign indicates that the 
large bracket in Eq.~(\ref{mel}) is approximated 
by 2, with increasing accuracy for increasing 
$|\vec p|$. The result is infinite. 

The consequence is that one does not know 
how to calculate the ground state of the 
interacting theory, or vacuum, in terms 
of the ground state of a free theory. All 
excited states contain similar divergences
as the vacuum state.

To avoid the infinities and attempt a search
for the spectrum of a regulated quantum 
Hamiltonian, one might impose a cutoff on 
$|\vec p|$. The price would be a violation 
of Lorentz symmetry, since boosts can change 
$|\vec p|$ by an arbitrary amount. 

Similar interaction terms appear in QED. 
For example, there are terms in the QED 
IF Hamiltonian that are capable of creating 
the electron, positron, and photon out of 
bare vacuum and thus produce divergences
in a similar pattern. These interaction terms 
and the resulting effects in states containing
electrons and photons, are known how to handle 
in perturbative calculations~\cite{Feynman R(1949)}. 
But beyond the perturbative rules, one may 
question the existence of QED in the IF 
Schr\"odinger picture~\cite{Dirac P(1965)}. 
In the case of QCD, in the domain where 
perturbation theory does not apply, the role 
of interaction terms that alter the bare vacuum 
in the IF approach, is not understood yet.

In summary, the mass mixing model provides
an elementary example of how the vacuum issue
arises in relativistic theories of physical 
interest, although in the realistic theories 
the relevant interaction terms are much more 
complicated than in the model. As a consequence, 
the general vacuum issue remains unsolved. 
Concerning the physics of the vacuum in the 
context of comparison between the IF and FF 
approaches, especially in the context of QCD,
see Refs.~\cite{Wilson et al.(1994),
Brodsky et al.(2012)} and the rich literature 
quoted there. 

\section{ IF re-quantization approach to 
the mass mixing interaction and vacuum issue }

In the IF approach, when one faces 
the vacuum problem due to the mass mixing
interaction terms, one can step back to the 
Lagrangian density and re-write it using
new field variables. Namely, one can write
the density using a $2 \times 2$-matrix 
notation, in which the fields $\psi$ and
$\phi$ form a doublet and a mass-matrix, say 
$M$, is introduced;
\beq
\label{cLPsi}
\cL  \es \bar \Psi ( i \partial \hspace{-5.5pt}/  - M) \Psi 
\quad , \quad \quad
\Psi \rs \left[ \begin{array}{c} \psi \\ 
\phi \end{array} \right] 
\quad , \quad 
M    \rs \left[ \begin{array}{cc} ~\mu~~ & m~~ \\ 
                                  ~m~~ & \nu~~ \end{array} \right] 
\, .
\eeq
The eigenvalues and corresponding eigenvectors 
of $M$ are
\beq
\label{m12}
m_{1,2}
\es \left[ \, \mu + \nu  \, \pm \, 
             (\mu - \nu) \, \epsilon \, \right]/2 \, , \\
\label{v12}
v_1 \es \left[ \begin{array}{r} \cos \varphi  \\
                              - \sin \varphi
               \end{array} \right]  
\quad , \quad 
v_2 \rs \left[ \begin{array}{r} \sin \varphi  \\
                                \cos \varphi
               \end{array} \right] \, ,
\eeq 
where $ \epsilon = \sqrt{1 + [2 m/(\mu-\nu)]^2} $
and $ \varphi = - \arctan{ \sqrt{ (\epsilon - 1)/
                                  (\epsilon + 1) } }$.
One can write the doublet field $\Psi$ in terms
of the eigenvectors, $\Psi = \psi_1 \, v_1 + \psi_2 \, v_2$,
and use the fields 
\beq
\label{xi1}
\psi_1 \es \cos \varphi \,\, \psi -  \sin \varphi \,\, \phi 
\quad {\rm and } \quad  
\psi_2 \rs \sin \varphi \,\, \psi +  \cos \varphi
\,\, \phi  
\eeq
as new degrees of freedom in a classical theory.
In terms of the new fields, the Lagrangian density
reads
\beq
\label{Lreq}
\cL 
\es 
\bar \psi_1 (i \partial \hspace{-5.5pt}/ - m_1) \psi_1 
+
\bar \psi_2 (i \partial \hspace{-5.5pt}/ - m_2)
\psi_2 \, ,
\eeq
which is a theory of free fields corresponding
to masses $m_1$ and $m_2$. 

Section~\ref{Qfree} explains that the terms of type 
$\hat h$ in Eq.~(\ref{hIF}) do not appear in the 
corresponding IF quantum Hamiltonian once the 
right energies are used in the construction of 
quantum field operators, including boosted spinors 
and corresponding creation and annihilation operators. 

The result is that one can avoid the vacuum issue 
in quantum theory when one constructs it by 
quantizing the field degrees of freedom that 
appear in Eq.~(\ref{Lreq}) rather than those 
that appear in Eq.~(\ref{LIFI}). Since one 
steps back from a divergent quantum theory 
with interaction to the initial classical 
one and then quantizes the classical theory 
again using new degrees of freedom, for which 
the troublesome divergence turns out to cancel 
out, the IF cure for the vacuum problem can be 
called re-quantization.
 
In the re-quantization, one uses $E_{m_1 p}$ and 
$E_{m_2 p}$ instead of $E_{\mu p}$ and $E_{\nu p}$ 
in constructing quantum field operators $\hat 
\psi_1$ and $\psi_2$, with new creation and 
annihilation operators. The resulting IF 
Hamiltonian,
\beq
\label{reqHIFQ2}
\hat H \es \hspace{-11pt}
\sum_{~~~~~~ m_1 p s} \hspace{-25pt} \int 
\,
E_{m_1 p} \,
\left( b^\dagger_{m_1 p s}  b_{m_1 p s}
     + d^\dagger_{m_1 p s}  d_{m_1 p s} \right)
+
\hspace{-12pt}
\sum_{~~~~~~ m_2 p s} \hspace{-25pt} \int 
\,
E_{m_2 p} \,
\left( b^\dagger_{m_2 p s}  b_{m_2 p s}
     + d^\dagger_{m_2 p s}  d_{m_2 p s} \right) \, ,
\eeq
describes free fermions of masses $m_1$ and 
$m_2$ as desired in a relativistic theory.

The key question one faces in the IF of 
Hamiltonian dynamics is how to deal with 
the vacuum issue in the presence of complex
interactions, i.e., when one does not know 
what masses to use. This becomes a serious
roadblock when one does not even know if 
such masses exist. In case of theories such 
as QCD, which are expected to describe confined 
fermions, the very existence of ``right'' masses 
for IF quantization is doubtful.

\section{ FF canonical quantum theory for fermions with mass mixing interaction }

Previous consideration suggests that 
one needs some alternative approach 
to the IF re-quantization in order 
to handle theories with complex 
interactions. The elementary fermion 
model with mass mixing interaction is 
used here to explain how a candidate 
for such alternative method, the RGPEP, 
avoids the vacuum problem and produces 
the same solution that the IF re-quantization 
allows one to find. The FF approach to 
the model does not require any re-quantization. 
With the preliminaries laid out above, 
the description of FF approach given 
below is considerably abbreviated.

The space-time hyperplane, which in 
the FF of Hamiltonian dynamics is used 
instead of the hyperplane $t=x^0=0$, 
is defined by the condition $x^+=0$. 
The notation is $x^\pm = x^0 \pm x^3$ and
$x^\perp = (x^1,x^2)$, the same for all
tensor indices. Other conventions used
here are explained in Ref.~\cite{Glazek S(2013)}.

The Euler-Lagrange equation for the doublet 
fermion field in the Lagrangian density of 
Eq.~(\ref{cLPsi}), namely $( i \partial 
\hspace{-5.5pt}/  - M) \, \Psi = 0$, 
can be written in terms of two orthogonal 
components $\Psi_\pm = \Lambda_\pm \Psi$,
where $\Lambda_\pm = \gamma^0 \gamma^\pm/2$
are projection matrices, as an equation 
of motion and a constraint, respectively,
\beq
\label{projectedM2}
i \partial^- \Psi_+ 
\es 
( i \alpha^\perp \partial^\perp + \beta M) \Psi_- 
\quad {\rm and } \quad
i \partial^+ \Psi_- 
\rs 
( i \alpha^\perp \partial^\perp + \beta M) \Psi_+ \, .
\eeq
Following Refs.~\cite{Chang S(1973),Chang and Yan(1973)},
one obtains the FF Hamiltonian, 
\beq
P^- \es {1 \over 2 }\int dx^- d^2x^\perp \, \cT^{+-} \, ,
\eeq 
where the relevant component of the energy-momentum 
density tensor is
\beq
{ 1\over 2} \cT^{+-} 
\es \Psi_+^\dagger i \partial^- \Psi_+ 
\rs
\Psi_+^\dagger 
( i \alpha^\perp \partial^\perp + \beta M)  
{ 1 \over i \partial^+} \,  
( i \alpha^\perp \partial^\perp + \beta M) \Psi_+ \, .
\eeq
The resulting FF quantum Hamiltonian is
obtained by replacing $\Psi$ by the
quantum field operator $\hat \Psi$ and 
evaluating
\beq
\label{P-1}
\hat P^- \es \int dx^- d^2x^\perp \, 
: \hat \Psi_+^\dagger \, { - \partial^{\perp \, 2} + M^2  
\over i \partial^+} \,  \hat \Psi_+ : \ .
\eeq 
In the FF representation of 
$\gamma$-matrices~\cite{Glazek S(2013)}, the 
operator $\hat \Psi_+$ has the form 
\beq
\Psi_+(x) \rs \left[ \begin{array}{c} \hat \zeta(x) \\
                                      0        \\ 
                                     \hat \omega(x) \\
                                      0
\end{array}\right] \, ,
\eeq
where $\hat \zeta$ and $\hat \omega$ are two-component 
fields (the field $\Psi$ has eight components). The 
anti-commutation relations at $x^+=x'^+=0$ are set in 
the form
\beq
\label{crqftext}
\left\{ \hat \zeta(x),  \hat \zeta^\dagger(x') \right\}
\es
\left\{ \hat \omega(x), \hat \omega^\dagger(x') \right\}
\rs 
\delta^3(x - x') \, .
\eeq
The FF Fourier expansions of operators $\hat \zeta$ and
$\hat \omega$ are independent of mass parameters. Namely,  
\beq
\label{zetaqftext}
\hat \zeta(x) 
\es 
\sum_{~~ps} \hspace{-13pt}\int  \, \sqrt{p^+} \,
\left[  b_{\zeta ps}  \, e^{-ipx} 
      - d_{\zeta ps}^\dagger \, e^{ ipx} \sigma^1
\right] \, \chi_s  \, , \\
\label{omegaqftext}
\hat \omega(x) 
\es 
\sum_{~~ps} \hspace{-13pt}\int  \, \sqrt{p^+} \,
\left[  b_{\omega ps}  \, e^{-ipx} 
      - d_{\omega ps}^\dagger \, e^{ ipx} \sigma^1
\right] \, \chi_s  \, ,
\eeq
where 
\beq
\label{notationqft}
\sum_{~~ps} \hspace{-13pt}\int 
\es
\sum_{s = \, \pm 1} \int_{-\infty}^{+\infty}  { d^{2} p^\perp \over (2\pi)^2} \,
\int_0^{+\infty}  {d p^+ \over 2(2\pi) p^+} \, ,
\eeq
and $\chi_s$ is the two-component Pauli 
spinor normalized to 1 (the Pauli matrix 
$\sigma^1$ takes care of the required spin 
flip and the negative sign reflects the 
requirement of Fermi-Dirac statistics). 
The creation and annihilation operators 
satisfy mass-independent anti-commutation 
relations
\beq
\label{bdzeta}
\left\{ b_{\zeta ps}, b^\dagger_{\zeta p's'} \right\}
\es
\left\{ d_{\zeta ps}, d^\dagger_{\zeta p's'} \right\}
\rs
\left\{ b_{\omega ps}, b^\dagger_{\omega p's'} \right\}
\rs
\left\{ d_{\omega ps}, d^\dagger_{\omega p's'} \right\}
\rs
2p^+ (2\pi)^3 \delta^3(p - p') \, \delta_{s s'} \, .
\eeq
For comparison with the IF theory, it is important
to stress that the FF Fourier expansion of fields 
does not require any knowledge of mass or ``energy''
$p^-$ of the field quanta, in stark contrast to the 
IF construction where one assumes a value for the 
mass of field quanta and change of mass requires
re-quantization. No re-quantization is needed in
the FF version of the interacting theory.

Evaluation of the FF Hamiltonian yields
\beq
\label{P-bd}
\hat P^-
\es 
\sum_{~~ps} \hspace{-12pt}\int \,
  \left[ \left( p_\mu^- + { m^2 \over p^+} \right) 
  \,\left( b^\dagger_{\zeta p s} \, b_{\zeta p s} 
         + d^\dagger_{\zeta p s} \, d_{\zeta p s} \right)
+
         \left( p_\nu^- + { m^2 \over p^+} \right) 
  \,\left( b^\dagger_{\omega p s} \, b_{\omega p s} 
         + d^\dagger_{\omega p s} \, d_{\omega p s} \right)
\right.
\np
\left.
        { m(\mu + \nu) \over p^+} \,
  \, \left( b_{\zeta  p s}^\dagger \, b_{\omega p s}   
          + d_{\omega p s}^\dagger \, d_{\zeta  p s} 
+
            b_{\omega p s}^\dagger \, b_{\zeta  p s}   
          + d_{\zeta  p s}^\dagger \, d_{\omega p s} \right)
\right] \, ,
\eeq
where $p_\mu^- = (p^{\perp \, 2} + \mu^2)/p^+$ 
and   $p_\nu^- = (p^{\perp \, 2} + \nu^2)/p^+$.
The free fermion theory, the one before the mass
mixing terms are added, corresponds to the 
Hamiltonian denoted by $\hat P_f^-$, which is 
obtained from $\hat P^-$ by setting $m=0$. 

Note that in the FF Hamiltonian $\hat P^-$ 
there are no terms of the type $\hat h$ that 
cause trouble in the IF version of the 
theory. The result is that $\hat P^- 
|0\rangle = 0 $. Therefore, one may assume 
that the bare vacuum $|0\rangle$, which by 
definition is annihilated by all the annihilation
operators, can represent the physical vacuum.
This is precisely what happens in the model.
In more complex theories, one has to find 
out if the interactions allow for this 
interpretation of $|0\rangle$. For example, 
if very strong interactions produced eigenstates
of $\hat P^-$ with negative eigenvalues, one 
could not interpret $|0\rangle$ with eigenvalue 
zero as a ground state of a physical theory.
 
In FF Hamiltonians, the absence of interaction 
terms that can alter the bare vacuum state 
$|0\rangle$  is a generic feature of all 
theories of physical interest. This feature 
is a consequence of translation invariance, 
which implies conservation of $+$-component 
of total momentum in the dynamics. Since 
all quanta included in the theory have 
positive $p^+$, as indicated in the Fourier
spectrum of momenta in Eq.~(\ref{notationqft}),
the quanta cannot carry zero, which characterizes
the state $|0\rangle$. Thus, the translation
invariant Hamiltonian cannot create quanta from 
or annihilate them into the bare vacuum.

The only exception to stability of $|0\rangle$ 
is its interaction with the Fourier modes with 
$p^+=0$. But for quanta with non-zero masses such 
modes correspond to infinite eigenvalues of $\hat 
P^-_f$ and by definition these modes lie outside 
the cutoffs used to regulate a theory. If there 
is a need to include any effects due to such 
modes, the possibility which is not excluded 
in singular gauge theories, these effects should 
be identifiable through the cutoff dependence 
that appears when the needed modes are absent.
However, this means that their effects can be 
describable using counter terms to the cutoff 
dependence. This is explained Ref.~\cite{Wilson et al.(1994)}.
The cutoff dependence and counter terms can be 
studied using the RGPEP. The FF of dynamics 
thus appears to provide new ways for handling 
the vacuum problem that is otherwise prohibitively 
divergent in the IF of quantum dynamics. In the 
fermion model with mass mixing, no cutoff 
dependence arises that would require questioning 
$|0\rangle$ as a good vacuum state. 

In the model example, with terms of type $\hat h$ 
being absent for reasons found in all physically 
interesting theories with regularization, the FF 
Hamiltonian still contains interaction terms that 
mix fermions of masses $\mu$ and $\nu$ with arbitrary
strength. The question is how to find the spectrum. 
This is done below using the RGPEP within one and 
the same quantum theory, no re-quantization being
involved.

\section{ The RGPEP in quantum field theory }

General non-perturbative definition of the 
RGPEP in quantum field theory can be found
in Ref.~\cite{Glazek S(2012)}. In its essence,
the procedure amounts to transforming quantum
fields from the bare ones to effective ones
by a unitary rotation, $ \psi_t = \cU_t \, 
\psi_0 \, \cU_t^\dagger $. This is done by
expressing the bare, point-like quanta of a
local theory in the canonical Hamiltonian 
with counter terms by the quanta of size
$s = t^{1/4}$. As a result, one obtains a 
non-local theory of quanta that are called 
effective particles. So, for fermions, $ 
b_{t p} = \cU_t \, b_{0 p} \, \cU^\dagger_t $ and
$ d_{t p} = \cU_t \, d_{0 p} \, \cU^\dagger_t $.
The Hamiltonian remains the same,
$ \cH_t(b_t, d_t) = \cH_0(b_0, d_0) $, but 
it is written using different degrees of 
freedom. Below, where necessary for brevity, 
the annihilation operators $b$ and $d$ are 
commonly denoted by $q$.

The canonical theory with counter terms has 
a Hamiltonian of the form
\beq
\cH_0(q_0) \es
\sum_{n=2}^\infty \, 
\sum_{i_1, i_2, ..., i_n} \, c_0(i_1,...,i_n) \, \, q^\dagger_{0i_1}
\cdot \cdot \cdot q_{0i_n} \, .
\eeq
The effective theory Hamiltonian differs by 
a change of operators $q_0$ to $q_t$ and
coefficients $c_0$ to $c_t$,
\beq
\cH_t(q_t) \es
\sum_{n=2}^\infty \, 
\sum_{i_1, i_2, ..., i_n} \, c_t(i_1,...,i_n) \, \, q^\dagger_{ti_1}
\cdot \cdot \cdot q_{ti_n} \, .
\eeq
The goal of RGPEP is to find the coefficients
$c_t$. This is done by defining $\cH_t(q_0) = 
\cU^\dagger_t \, \cH_0(q_0) \, \cU_t$ and 
solving the equation 
\beq
\cH'_t(q_0) \es [ \cG_t(q_0) , \cH_t(q_0) ] \, ,
\eeq
where prime denotes the derivative with respect to $t$
and $\cG_t = - \cU_t^\dagger \cU'_t$. Knowing
$\cG_t$, one can find
\beq
\cU_t 
\es
T \exp{ \left( - \int_0^t d\tau \, \cG_\tau
\right) } \, ,
\eeq 
where $T$ denotes ordering operators 
according to the size of effective 
particles. In its simplest version, 
the RGPEP defines $\cG_t$ by the formula
\beq
\cG_t \es [ \cH_f, \cH_{Pt}]  \, , 
\eeq
where $\cH_f$ is the free part of the initial
Hamiltonian and  
\beq
\cH_{Pt}(q_0) \es
\sum_{n=2}^\infty \, 
\sum_{i_1, i_2, ..., i_n} \, c_t(i_1,...,i_n) \, \,\left( {1 \over
2}\sum_{k=1}^n p_{i_k}^+ \right)^2 \, \, q^\dagger_{0i_1}
\cdot \cdot \cdot q_{0i_n} \, ,
\eeq
which means that $\cH_{P t}$ differs from $\cH_t$
only by multiplication of the coefficients $c_t$
by a square of the total $+\,$-momentum carried by 
quanta involved in a term.

In summary, the simplest version of RGPEP is carried 
out by solving the equation 
\beq
\cH'_t \es \left[ [ \cH_f, \cH_{Pt} ], \cH_t \right] 
\eeq
with initial condition $\cH_0 \rs \cH_{canonical}
+ \cH_{CT} $, where $\cH_{CT}$ denotes the
counter terms. The counter terms are also found
using the RGPEP; one adjusts the initial
conditions so that the effective theories match
physical data by their solutions.

\section{ Application of the RGPEP to the fermion mass-mixing model }

The starting point is the Hamiltonian of 
Eq.~(\ref{P-bd}), which is treated as an
initial condition for a solution to 
\beq
\label{RGPEPeq}
{\cP^-_t}' 
\es
\left[ [ \cP^-_f, \cP^-_{Pt} ], \cP^-_t \right] \, .
\eeq 
A great simplification occurs in the fermion 
model: the initial condition does not require 
counter terms that diverge as functions of 
regularization parameters when regularization 
is being removed. The model does not require 
regularization of divergences; only finite 
adjustments of initial masses $\mu$ and $\nu$ 
and mass mixing parameter $m$ are required in 
order to obtain desired values of fermion 
masses in the resulting spectrum.

By examining the RGPEP equations in the model,
one finds that the solution for $\cP^-_t$
has the form 
\beq
\label{Pt}
\cP^-_t 
\es
\sum_{~~ps} \hspace{-13pt}\int \,
  \left[ A_{tp} 
  \, \left( b^\dagger_{\zeta p s} \, b_{\zeta p s} 
          + d^\dagger_{\zeta p s} \, d_{\zeta p s} \right)
+
         B_{tp}  
  \, \left( b^\dagger_{\omega p s} \, b_{\omega p s} 
          + d^\dagger_{\omega p s} \, d_{\omega p s} \right) 
\right.
\np
\left.
        C_{tp} 
  \, \left( b_{\zeta  p s}^\dagger \, b_{\omega p s}    
          + b_{\omega p s}^\dagger \, b_{\zeta  p s}
+
            d_{\zeta  p s}^\dagger \, d_{\omega p s}  
          + d_{\omega p s}^\dagger \, d_{\zeta  p s} \right)
\right] \, ,
\eeq
where the coefficients can be written in the forms
\beq
\label{At}
A_{tp}    \es { p^{\perp \, 2} + \mu_t^2 \over p^+ } 
\quad \quad , \quad \quad 
B_{tp}    \rs { p^{\perp \, 2} + \nu_t^2 \over p^+ } 
\quad \quad , \quad \quad 
C_{tp}    \rs {                    m_t^2 \over p^+
} \, ,
\eeq
and the initial conditions for the mass parameters 
are
\beq
\label{mu0}
\mu_0^2 \es \mu^2 + m^2 
\quad \quad , \quad \quad 
\nu_0^2 \rs \nu^2 + m^2 
\quad \quad , \quad \quad 
m_0^2   \rs m(\mu+\nu) \, .
\eeq
The RGPEP generator in Eq.~(\ref{RGPEPeq}) involves
\beq
\cP_f^-
\es 
\sum_{~~ps} \hspace{-13pt}\int \,
  \left[   p^-_\mu
  \,\left( b^\dagger_{\zeta p s} \, b_{\zeta p s} 
         + d^\dagger_{\zeta p s} \, d_{\zeta p s} \right)
+
           p^-_\nu
  \,\left( b^\dagger_{\omega p s} \, b_{\omega p s} 
         + d^\dagger_{\omega p s} \, d_{\omega p
s} \right) \right] \, , \\
\cP^-_{Pt} 
\es
\sum_{~~ps} \hspace{-13pt}\int \, p^{+ \, 2}\,
  \left[ A_{tp} 
  \, \left( b^\dagger_{\zeta p s} \, b_{\zeta p s} 
          + d^\dagger_{\zeta p s} \, d_{\zeta p s} \right)
+
         B_{tp}  
  \, \left( b^\dagger_{\omega p s} \, b_{\omega p s} 
          + d^\dagger_{\omega p s} \, d_{\omega p s} \right) 
\right.
\np
\left.
        C_{tp}  
     \left( b_{\zeta  p s}^\dagger \, b_{\omega p s}    
          + b_{\omega p s}^\dagger \, b_{\zeta  p s}
+
            d_{\zeta  p s}^\dagger \, d_{\omega p s}  
          + d_{\omega p s}^\dagger \, d_{\zeta  p s} \right)
 \right] \, ,
\eeq
and hence has the form
\beq
\label{generator}
{[} \cP_f^-, \cP_{Pt}^- {]}
\es
\sum_{~~ps} \hspace{-13pt}\int \, 
        C_{tp} \, p^{+ \, 2}(p_\mu^- - p_\nu^- )
\,
 \left( b_{\zeta  p s}^\dagger \, b_{\omega p s}
      - b_{\omega p s}^\dagger \, b_{\zeta  p s}
      + d_{\zeta  p s}^\dagger \, d_{\omega p s}  
      - d_{\omega p s}^\dagger \, d_{\zeta  p s}
\right) \, .
\eeq
By equating coefficients in front of the same 
products of creation and annihilation operators 
on both sides of Eq.~(\ref{RGPEPeq}), one obtains
three spin-independent equations for every value 
of momentum $p$, 
\beq
A'_{tp} 
\es
  2 p^{+ \, 2} \, (p^-_\mu - p^-_\nu) \, C^2_{tp}
\, , \\
B\hspace{1pt}'_{tp} 
\es
- 2 p^{+ \, 2} \, (p^-_\mu - p^-_\nu) \, C^2_{tp}
\, , \\
C\hspace{1pt}'_{tp} 
\es 
-   p^{+ \, 2} \, (p^-_\mu - p^-_\nu) \, 
(A_{tp} - B_{tp}) \, C_{tp} \, .
\eeq
But when one inserts in this set the coefficients 
$A_t$, $B_t$ and $C_t$ in their forms shown in
Eqs.~(\ref{At}), the momenta $p^\perp$ and $p^+$ 
drop out and all these RGPEP equations reduce to 
just three for the mass parameters,
\beq
\label{mut}
\left( \mu_t^2 \right)'
\es
  2 \, \left( \mu^2 - \nu^2 \right) \, \left(
m_t^2 \right)^2 \, , \\
\label{nut}
\left( \nu_t^2 \right)'
\es
- 2 \, \left( \mu^2 - \nu^2 \right) \, \left(
m_t^2 \right)^2 \, , \\
\label{mt}
\left( m_t^2 \right)'
\es 
- \left( \mu  ^2 - \nu  ^2 \right) 
\, 
\left( \mu_t^2 - \nu_t^2 \right) \, m_t^2 \, .
\eeq
This simplification has to happen because the 
RGPEP Eq.~(\ref{RGPEPeq}) is designed to respect 
the seven kinematical Poincar\'e symmetries of 
the FF of Hamiltonian dynamics. The three 
equations for mass parameters can be written
in the form of one matrix equation,
\beq
\left( \begin{array}{cc}
               \mu_t^2  &    m_t^2      \\
                 m_t^2  &  \nu_t^2
               \end{array} \right)'
\es
\left[
\left[  \left( \begin{array}{cc}
               \mu^2  &  0            \\
               0      &  \nu^2
               \end{array} \right),
\left( \begin{array}{cc}
               0      &   m_t^2       \\
               m_t^2  &   0
               \end{array} \right) \right],
\left( \begin{array}{cc}
               \mu_t^2 &  m_t^2      \\
                m_t^2  &  \nu_t^2
               \end{array} \right)
\right] \, ,
\eeq
which would be Wegner's flow equation~\cite{Wegner F(1994)}
for a $2 \times 2$ Hamiltonian matrix, if the 
first, diagonal matrix on the right-hand 
side contained $\mu_t^2$ and $\nu_t^2$
instead of $\mu^2$ and $\nu^2$, respectively.
Solutions are
\beq
\label{mutA}
\mu_t^2 \es m^2 + {1 \over 2} \, (\mu^2 + \nu^2) +
{ 1 \over 2} \, \delta \mu^2_t \, , \\
\label{nutA}
\nu_t^2 \es m^2 + {1 \over 2} \, (\mu^2 + \nu^2) -
{ 1 \over 2} \, \delta \mu^2_t \, , \\
\label{deltamutA}
\delta \mu^2_t  \es     (\mu^2 - \nu^2) \, 
                      { \cosh { x_t } + \epsilon \sinh { x_t }
                \over
                        \cosh { x_t } + \epsilon^{-
1} \sinh { x_t } } \, , \\
\label{mtA}
        m_t^2   \es {  m (\mu+\nu)  \over
                        \cosh { x_t } +
\epsilon^{-1} \sinh { x_t } } \, , 
\eeq
where $x_t = (\mu^2-\nu^2)^2 \, \epsilon  \, t$
and $\epsilon$ is introduced in Eq.~(\ref{m12}).

It is visible that $m_t^2$ tends to zero when $t
\rightarrow \infty$. As a result, one obtains
in the limit a theory of two types of free fermions 
that do not mix with each other. Masses squared of 
these fermions are 
$\lim_{t \rightarrow \infty} \mu_t^2 = m_1^2$ 
and 
$\lim_{t \rightarrow \infty} \nu_t^2 = m_2^2$,
where $m_1$ and $m_2$ are the same as the 
IF re-quantization masses in Eq.~(\ref{m12}).
This is how the RGPEP recovers the IF result
in a quantum theory with empty vacuum and 
without re-quantization. The spectrum of solutions 
to the eigenvalue problem of FF Hamiltonian found 
using the RGPEP, consists of states of free fermions 
with masses $m_1$ and $m_2$ in empty vacuum. The
solutions form a Poincar\'e invariant spectrum
in as large a range of momenta as one wishes.

\section{ Conclusion }

The model discussion can be concluded by observing 
that fermions cannot become tachyons. The issue arises
because the free FF Hamiltonian depends on the squares
of masses of fermions. A priori, one could expect
that, in the presence of mass-mixing interactions, 
the eigenvalues of the matrix of masses squared could 
have negative eigenvalues, and thus describe tachyonic 
fermions. However, the RGPEP does not allow this to 
happen. One can write 
\beq
\lim_{ t \rightarrow \infty }
\left( 
\begin{array}{cc} 
\mu_t^2   &  m_t^2 \\
m_t^2     &  \nu_t^2
\end{array}
\right)
\es
\lim_{ t \rightarrow \infty }
\left(
\begin{array}{cc} 
m_{1t} & m_{It} \\
m_{It} & m_{2t}
\end{array}
\right)^2
\rs
\left(
\begin{array}{cc} 
m_1^2 & 0     \\
0     & m_2^2
\end{array}
\right) \, ,
\eeq
and see that the squares of masses cannot become
negative in the fermion model since the eigenvalues 
$m_1$ and $m_2$ are real.

In the model, the RGPEP avoids the vacuum 
problem and provides information about what 
happens when the mass mixing interaction terms 
are so large, $|m| > \sqrt{ \mu \nu}$, that 
the smaller one of the two IF mass eigenvalues 
becomes negative. Namely, a chiral rotation can 
restore the positive sign of the eigenmass,
while the RGPEP mass squared never approaches 
zero when $t$ changes from zero to infinity. 
Instead, $\mu_t^2$ and $\nu_t^2$ go directly to 
the positive squares of the eigenmasses $m_1$ 
and $m_2$ irrespective of the signs of $m_1$ 
and $m_2$. More detailed explanation of this 
result is available in~\cite{Glazek S(2013)}. 

The elementary model of fermions with mass mixing 
interaction illustrates the RGPEP as a new tool 
for studying quantum field theory. The method
passes a test of providing solutions one expects 
to be valid on the basis of the IF re-quantization 
approach. It is important to observe that nowhere 
in the RGPEP any assumption was made to the effect 
that the interactions are small; no perturbative 
expansion was employed and the theory is solved 
using the RGPEP for arbitrary strength of the 
interaction terms. Still, in complex theories where
non-perturbative calculations require considerable
gain in understanding what actually happens and
perturbative studies may provide helpful signposts, 
one may also apply the RGPEP using a perturbative 
expansion~\cite{Glazek S(2012)}.   

One may conclude by quoting Dirac~\cite{Dirac P(1978)}, 
who said that the front form ``offers new opportunities, 
while the familiar instant form seems to be played out.''


\begin{thebibliography}{3}

\bibitem[Glazek S(2013)]{Glazek S(2013)}
G{\l}azek S D (2013)
Fermion mass mixing and vacuum triviality in 
the renormalization group procedure for effective particles.
Phys. Rev. D 87: 125032--12

\bibitem[Dirac P(1949)]{Dirac P(1949)}
Dirac P A M (1949) 
Forms of Relativistic Dynamics.
Rev. Mod. Phys. 21: 392--399

\bibitem[Dirac P(1965)]{Dirac P(1965)}
Dirac P A M (1965) 
Quantum Electrodynamics without Dead Wood.
Phys. Rev. 139: B684--B690 

\bibitem[Wigner E(1938)]{Wigner E(1938)}
Wigner E P (1938) 
On Unitary Representations of the Inhomogeneous Lorentz Group.
Ann. of Math. 40: 149--204

\bibitem[Heisenberg and Pauli(1929)]{Heisenberg and Pauli(1929)}
Heisenberg W, Pauli W (1929)
Zur Quantentheorie der Wellenfelder. 
Z. f. Phys. 56: 1--61 

\bibitem[Heisenberg and Pauli(1930)]{Heisenberg and Pauli(1930)}
Heisenberg W, Pauli W (1930)
Zur Quantentheorie der Wellenfelder. II.
Z. f. Phys. 59: 168--190 

\bibitem[Bjorken and Drell(1964)]{Bjorken and Drell(1964)}
Bjorken J D, Drell S D (1964) 
Relativistic Quantum Mechanics.
McGraw-Hill, New York

\bibitem[Bjorken and Drell(1965)]{Bjorken and Drell(1965)}
Bjorken J D, Drell S D (1965) 
Relativistic Quantum Fields.
McGraw-Hill, New York

\bibitem[Weinberg S(1995)]{Weinberg S(1995)}
Weinberg S (1995)
The Quantum Theory of Fields.
Cambridge University Press, Cambridge 

\bibitem[Peskin M(1995)]{Peskin M(1995)}
Peskin M E, Schroeder D V (1995)
An Introduction to Quantum Field Theory.
Addison-Wesley, Reading

\bibitem[Wilson et al.(1994)]{Wilson et al.(1994)}
Wilson K G et al. (1994)
Nonperturbative QCD: A weak-coupling treatment on the light front.
Phys. Rev. D 49: 6720--6766

\bibitem[Brodsky et al.(2012)]{Brodsky et al.(2012)}
Brodsky S J,  Roberts C D, Shrock R, Tandy P C (2012)
Confinement contains condensates.
Phys. Rev. C 85: 065202--9

\bibitem[Feynman R(1949)]{Feynman R(1949)}
Feynman R P (1949)
Space-Time Approach to Quantum Electrodynamics.
Phys. Rev. 76: 769--789

\bibitem[Chang S(1973)]{Chang S(1973)}
Chang S J, Root R G, Yan T M (1973) 
Quantum Field Theories in the Infinite-Momentum Frame. I. 
Quantization of Scalar and Dirac Fields.
Phys. Rev. D 7: 1133-1146

\bibitem[Chang and Yan(1973)]{Chang and Yan(1973)}
Chang S J, Yan T M (1973) 
Quantum Field Theories in the Infinite-Momentum Frame. II. 
Scattering Matrices of Scalar and Dirac Fields.
Phys. Rev. D 7: 1147-1161

\bibitem[Glazek S(2012)]{Glazek S(2012)}      
G{\l}azek S D (2012)
Perturbative formulae for relativistic interactions
of effective particles. 
Acta Phys. Pol. B 43: 1843--1862

\bibitem[Wegner F(1994)]{Wegner F(1994)}
Wegner F (1994)
Flow Equations for Hamiltonians.
Ann. Phys. (Berlin) 3: 77--91

\bibitem[Dirac P(1978)]{Dirac P(1978)}
Dirac P A M (1978) 
The Mathematical Foundations of Quantum Theory.
In: Marlow A R (ed.)
Mathematical Foundations of Quantum Theory,
Loyola University, New Orleans, June 2-4, 1977, 
pp. 1-8. Academic Press, New York

\end{thebibliography}
\end{document}